# Sensitive Data Detection with High-Throughput Machine Learning Models in Electrical Health Records

Kai Zhang, PhD, Xiaoqian Jiang, PhD

The University of Texas Health Science Center,
McWilliams School of Biomedical Informatics, Houston, TX, USA

**Abstract:**
In the era of big data, there is an increasing need for healthcare providers, communities, and researchers to share data and collaborate to improve health outcomes, generate valuable insights, and advance research. The Health Insurance Portability and Accountability Act of 1996 (HIPAA) is a federal law designed to protect sensitive health information by defining regulations for protected health information (PHI). However, it does not provide efficient tools for detecting or removing PHI before data sharing. One of the challenges in this area of research is the heterogeneous nature of PHI fields in data across different parties. This variability makes rule-based sensitive variable identification systems that work on one database fail on another. To address this issue, our paper explores the use of machine learning algorithms to identify sensitive variables in structured data, thus facilitating the de-identification process. We made a key observation that the distributions of metadata of PHI fields and non-PHI fields are very different. Based on this novel finding, we engineered over 30 features from the metadata of the original features and used machine learning to build classification models to automatically identify PHI fields in structured Electronic Health Record (EHR) data. We trained the model on a variety of large EHR databases from different data sources and found that our algorithm achieves 99% accuracy when detecting PHI-related fields for unseen datasets. The implications of our study are significant and can benefit industries that handle sensitive data.

1. Introduction

Because of improvements in online data tracking and sharing techniques, healthcare data privacy has become a major issue in recent years. In the fields of medicine and research, the sharing of data with other parties and for secondary uses is a frequent practice [1]. Patients, however, have voiced worries regarding the lack of control they have over how their data is used and shared [2]. There are many difficulties in maintaining data confidentiality in the context of health care research[3]. Although there are legal ways to share data, there is always room for improvement in terms of speeding up and securing data transmissions. To encourage the safe and responsible use of healthcare data, it is critical to address these concerns as soon as they arise.

De-identification is the process of taking personal data out of a dataset so that it cannot be connected to particular people[4]. This procedure is crucial for safeguarding the privacy of people whose personal data is present in datasets. Personal information is described as data that can be used to identify an individual[5]. In the healthcare sector, where patient data sets frequently contain sensitive information, de-identification is especially crucial[6]. The Health Insurance Portability and Accountability Act's (HIPAA) Privacy Regulation offers advice on the best ways to achieve de-identification while still adhering to HIPAA rules[7]. De-identifying datasets requires considering both direct and indirect identifiers. In contrast to indirect identifiers, which may not be sufficient on their own to identify an individual but can result in identification when paired with other information, direct identifiers are sufficient alone to potentially identify an individual [8]. For the protection of structured data, a variety of techniques and resources are available. They include cryptographic techniques that can safeguard patient data while upholding HIPAA compliance and preserving data links [9]. Another method for protecting privacy and promoting open data sharing is statistical data de-identification [10,11].

There are two primary methods for de-identification, expert determination, and safe harbor methods. Expert determination is a case-by-case approach where an expert determines if the data has been properly de-identified. In contrast, the safe harbor method involves removing specific identifiers from data, and if those identifiers have been removed, the data is considered de-identified [12]. One common de-identification mechanism is redaction, which involves erasing or expunging sensitive data from a record. These techniques include removing direct identifiers such as addresses and phone numbers, removing and re-coding specific fields, and using codes, algorithms, or pseudonyms to render individual records distinguishable. Another mechanism is generalizing, also known as k-anonymization, which involves grouping data in a way that ensures that individuals cannot be re-identified [13]. However, both approaches require the accurate identification of sensitive attributes from data to apply their protections.

While some organizations may choose to de-identify data sets manually, several software tools are available to aid in the process. The most relevant work includes tools such as ARX, which offer methods for evaluating privacy risks associated with data anonymization [14]. These tools are particularly useful when dealing with sensitive data. ARX is a desktop tool that evaluates the risk of re-identification of a dataset and offers a balance between privacy and data utility. Additionally, other tools such as sdcMicro, Amnesia, μ-Argus, and Privacy Analytics can be used for privacy risk evaluation and data anonymization [15,16]. However, it is important to note that these tools do not focus on HIPAA-specific Personal Health Information (PHI). Therefore, organizations dealing with healthcare data still need to ensure they comply with the HIPAA regulations when de-identifying their data sets.

We would like to address this gap by providing solutions to identify PHIs from structured healthcare data using machine learning. One idea for training a machine learning model is to use the sampled values of each feature in a certain dataset as the input vector and use a machine learning model to predict whether this feature is a sensitive attribute. Alternatively, one can use deep learning models to learn embeddings of such vectors and use them for classification. Both theories imply that an attribute's data distribution can indicate if it is a sensitive attribute, and AI models are trained to recognize these associations. The data distribution of a characteristic can change significantly among datasets with diverse populations, making this technique problematic in some cases. As a result, the model's ability to generalize to other datasets is hampered. The applied test dataset's sensitive attribute columns must have similar distributions to the dataset the model was trained on for the model to predict sensitive attributes with the same level of accuracy. In actual circumstances, this is an extremely stringent requirement.

In this paper, we propose a novel approach to detecting PHI. Our approach is based on machine learning models, and we will model the sensitive attribute identification problem as a classification problem. Our approach involves extracting distributions and various statistics from the metadata of each field in the original dataset. According to a key finding from several datasets, sensitive properties frequently share characteristics in their metadata, despite the fact that their data distributions can differ between two real-world datasets. For example, certain sensitive attributes are always unique (IDs, medical record numbers (MRN), etc.), regardless of the amount of data samples in the database. It is very likely that attributes with fixed digits or strings are zip codes or IDs. Age is one example of an attribute that often falls within a specific range (e.g., 0-100). Even while field value distributions can vary for distinct datasets, such metadata features of a field in the EHR are frequently consistent across datasets. Based on this finding, we meticulously built 37 features from each field's information to use as inputs for the machine learning predictive model. This strategy can increase the efficiency of recognizing PHIs and acts as an alternative to the conventional de-identification procedure. By employing this technique, we want to speed up the de-identification procedure, guarantee the privacy of sensitive data, and allow for the continued use of the data for research and other uses. The difficulties of finding PHIs in structured data are addressed by our suggested method, which has the potential to be applied in a variety of contexts. Below is a succinct summary of our innovation:

1. We propose a novel approach to detecting Personal Health Information (PHI) from structured healthcare data based on machine learning models, which models the sensitive attribute identification problem as a classification problem.
2. We observe that the metadata of sensitive attributes often has similar characteristics across different datasets, and we designed 37 useful features from the metadata for PHI detection which has high generalization ability to different datasets with different populations.
3. We provide a unique perspective of viewing the sensitive attribute identification problem and provides an open-source toolkit based on the proposed idea that could facilitate research requiring sharing medical data.

**2. Methodology**

Sensitive attributes in the healthcare environment refer to sensitive patient health data that can identify a patient's identity. The identification can be direct (such as account numbers, medical record numbers, etc.) or indirect (such as name + date of birth). In this study, we use the "Safe Harbor" method in accordance with the Health Insurance Portability and Accountability Act of 1996 (HIPAA) Privacy Rule[7].

**2.1 PHI attributes**

The HIPAA's safe harbor method defined 18 categories of protected health information that can be used to infer the identity of an individual from healthcare data. Table 1 outlines the elements considered identifiable information under the HIPAA Privacy Rule in the healthcare industry. The data elements in this table consist of a range of personal identifying information, including names, social security numbers, dates of birth, and medical record numbers, along with other unique identifiers like biometric data, web URLs, and full-face photographs.

The major difficulty of a sensitive attribute identification system is that different datasets are very likely to use different formats or even encoding the plain text to numbers which loses semantic information (such as encoding White/Black/Asian to 0/1/2/…). In this case, regular expression methods usually fail. However, certain information is still preserved during this encoding process, such as the data type (categorical), number of unique categories, etc. In the following section, we will leverage this observation and introduce the two proposed methods.

*Table 1. List of Identifiable Data Elements in Healthcare Under HIPAA Privacy Rule*

| Category | Data Elements | Category | Data Elements |
|---|---|---|---|
| (A) Names | First and last name, middle initial, suffix | (J) Account numbers | Account number |
| (B) Geographic subdivisions | Street address, city, county, precinct, ZIP code | (K) Certificate/license numbers | Certificate/license number |
| (C) Dates | birth date, admission date, discharge date, death date, all ages > 89 and all elements of dates (including year) indicative of such age | (L) Vehicle identifiers and serial numbers | Vehicle identifier number, license plate number |
| (D) Telephone numbers | Area code, phone number | (M) Device identifiers and serial numbers | Device identifier number, serial number |
| (E) Fax numbers | Area code, fax number | (N) Web URLs | URL |
| (F) Email addresses | Email address | (O) Internet Protocol (IP) addresses | IP address |
| (G) Social security numbers | Social security number | (P) Biometric identifiers | Fingerprints, voiceprints |
| (H) Medical record numbers | Medical record number | (Q) Full-face photographs | Photographs |
| (I) Health plan beneficiary numbers | Health plan beneficiary number | (R) Other unique identifiers | Any other unique identifying number, characteristic, or code, except as permitted by paragraph (c) of this section |

**2.2 Methods**

There are two approaches to solving this problem: rule-based techniques and machine learning prediction models. We propose to build an ensemble pipeline by leveraging the advantages of the two methods to identify the sensitive attribute (Figure 1). In the following sections, we exchangeably use sensitive attribute or PHI to refer to the same meaning; we exchangeably use attribute or feature to refer to a column of the EHR dataset.

**2.2.1 Regular expression**

Our ensemble model is a powerful tool that consists of a rule-based module designed using regular expression. The primary objective of the rule-based screening tool is to identify sensitive attributes in Table 1, which often take different representations in different datasets. However, they mostly follow certain formats, and the number of formats for a particular field is often limited. This limitation provides us with a way to use regular expression pattern matching to identify them.

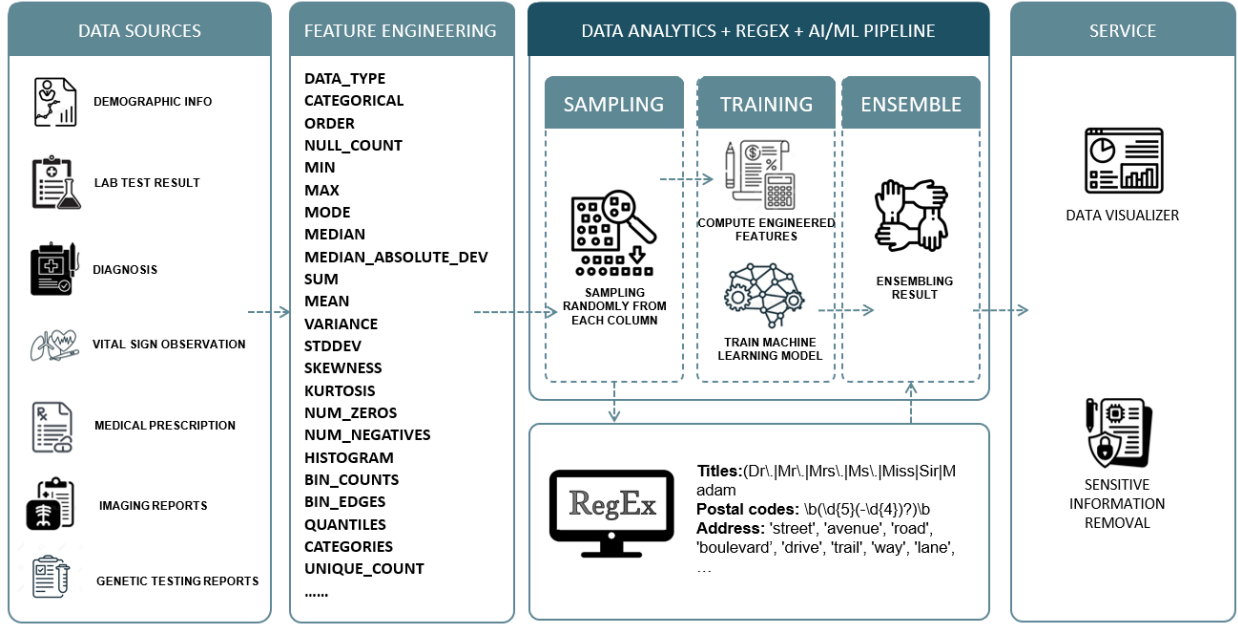

*Figure 1. Sensitive attribute identification pipeline*

For instance, a phone number can take various forms such as (123)456-7890, 1234567890, (123)4567890, 123-456-7890, and so on. By designing regular expression patterns that cover a wide range of all possibilities of the information belonging to the 18 categories in Table 1, we can accurately identify sensitive attributes and protect them from unauthorized access. Our rule-based module is based on the idea that sensitive attributes often follow specific patterns, and we can use regular expressions to identify them. This approach is highly effective in identifying sensitive information, even when it is represented differently in different datasets. By using regular expression pattern matching, we can quickly and accurately identify sensitive attributes and protect them from unauthorized access.

In order to identify sensitive attributes of the structured EHR data, we utilize a screening program that analyzes each attribute or feature column individually. To accomplish this, we randomly select data samples from each column and feed them into the screening program to screen for each of the common patterns we designed (see appendix). To determine the probability of each column being a sensitive attribute, we propose Algorithm 1, which is a strategic data sampling and screening process that yields the probability of each column being a sensitive attribute. This involves feeding the data and all patterns into the screening algorithm. Specifically, we randomly sample $K$ ($\leq N$) data points from a column ($m$) of the dataset $D$ (in total $M$ columns), where the value is not empty. We then perform regular expression matchings on each $K$ data point to determine the number of samples that fit into a certain pattern $R_p$. This allows us to calculate the probability of this column being a sensitive attribute. Once we have analyzed all the data samples within a given column, we can determine the maximum probability of this attribute ($m$) being any of the patterns $R_p$ (such as age, dob, address, MRN, etc.). This final probability is what we use to determine whether a given attribute contains private information. This same procedure is applied to all $M$ columns within the structured EHR data. By analyzing each attribute individually, we can identify sensitive attributes with a high degree of accuracy. This information can then be used to ensure that sensitive data is properly protected and secured, in order to maintain patient privacy and confidentiality.

**Algorithm 1:**
1 **Input**:
2     EHR tabular data $D$ ($N \times M$), Row: $N$ samples, Column: $M$ features, $[C_1, C_2, ..., C_M]$

```
   3        Regular expression patterns [R_1, ..., R_p] in Table 2
   4    Output:
   5        Probability of each column being a sensitive attribute
   6
   7    for m in [1, ..., M]:
   8    select K non-nan data samples at random from column C_m
   9        prob_PHI = 0
  10        for R_p in [R_1, ..., R_p]:
  11            num_match = ∑_{i∈K} I_{match(sample_i, R_p)}
  12            prob_PHI = max(prob_PHI, num_match / K )
  13        prob_PHI(m) = prob_PHI
  14    Return prob_PHI^{Regex}(C_1), ..., prob_PHI^{Regex}(C_M)
```

### 2.2.2 Machine learning

Sensitive attribute screening algorithms are essential in ensuring that personal information is protected. However, these algorithms can be problematic for two reasons. Firstly, it is usually difficult to exhaustively list all possible formats or representations of a sensitive attribute. This is because sensitive attributes can take on various forms, such as names, addresses, and social security numbers. Additionally, sensitive attributes can be subjective and context-dependent, making it challenging to identify them accurately. Secondly, sensitive attributes are not always shown in plain formats and can sometimes be coded. For instance, instead of using the traditional formats of (male/female, M/F, etc.) to indicate sex, some institutions may use 0/1. This can cause rule-based screening algorithms to fail, as they may not recognize the coded format.

To address this challenge, we propose an innovation in this study based on the observation that applying different coding strategies to the original data does not change its metadata. This means that even if institutions use different coding strategies to represent sensitive attributes, the resulting variables are still categorical with the same number of unique categories. This unique finding led the researchers to design a machine learning model to predict sensitive attributes based on engineered features of the metadata.

The machine learning model developed in this study contributes to a significant improvement over the rule-based screening algorithms. It can accurately predict sensitive attributes, even when they are coded using different strategies. This is because the model is designed to analyze the metadata of the data, rather than the data itself. By analyzing the metadata, the model can identify patterns and features that are indicative of sensitive attributes, regardless of how they are coded. See Algorithm 2.

**Feature engineering.** We manually engineered 30 features from the metadata of the columns that are most relevant to the prediction of whether an attribute contains sensitive information, Table 2. Note that even though the logic behind designing these engineered features is based on observations of various dataset, the logic is general but not always true, this is the reason why the engineered features are combined to be fed into the machine learning model to make predictions. The designing logic behind these features can also be found in the appendix.

*Table 2. Engineered features.*

| Data Type | Categorical | Order | Null Count | Min | Max | Mode | Median |
|---|---|---|---|---|---|---|---|
| Mean | Variance | Standard Deviation | Skewness | Kurtosis | Num Zeros | Num Negatives | Histogram |
| Quantiles | Categories | Unique Count | Unique Ratio | Categorical Count | Gini Impurity | Diversity Index | Precision |
| Median Absolute Deviation | Bin Edges | Bin Counts | Sum | | | | |

### 2.3 Ensembling

To build an ensemble model based on the probability predictions of the two methods, we need to calibrate the machine learning prediction probability first. The $M$ probabilities $prob_{PHI}^{ML}(C_1), ..., prob_{PHI}^{ML}(C_M)$ are fed into a logistic regression model to transform probabilities, which is also known as Platt scaling [17]. The scaled probabilities

$scaledprob_{PHI}^{ML}(C_1), \ldots, scaledprob_{PHI}^{ML}(C_M)$ will be ensemble with the regular expression result to get the final probability prediction for each feature $m$, i.e. $finalprob_{PHI}(C_m) = h(scaledprob_{PHI}^{ML}(C_m), prob_{PHI}^{Regex}(C_m))$. In our application, we choose the maximum function to be the function $h(\cdot,\cdot)$ due to our intent of not missing a sensitive attribute (false negative) but does not care much about wrongly classifying a non-sensitive attribute to sensitive attribute (false positives). In other words, false negatives are more important than false positives in this application.

---

**Algorithm 2:**
1. **Input**:
2.     EHR tabular data $D$ ($N \times M$), Row: $N$ samples, Column: $M$ features, $[C_1, C_2, \ldots, C_M]$
3.     Engineered feature calculation functions $F_1(\cdot), F_2(\cdot), \ldots, F_P(\cdot)$
4. **Output:**
5.     Probability of each column being a sensitive attribute
6. 
7. *for $m$ in $[1, \ldots, M]$:*
8.     *select K non-nan data samples at random from column $C_m$*
9.     *calculate P features for each column m: $F_1(D[1:N,1]), F_2(D[1:N,2]), \ldots$*
    *$F_P(D[1:N, M])$*
10. *Construct new dataset $D'$ ($M \times P$), where M is the original features, P are the engineered features*
11. *Feed $D'$ ($M \times P$) to the machine learning predictive models to predict the probability of each of the M features are sensitive attribute*
12. *Return $prob_{PHI}^{ML}(C_1), \ldots, prob_{PHI}^{ML}(C_M)$*

---

## 3. Results
### 3.1 Dataset selection and preprocessing

We collected eight EHR datasets from various data sources to cover a large variety of different formats of each sensitive attribute. We manually go over all the features in the above dataset to assign labels (0: non-sensitive attribute, 1: sensitive attribute) for each feature of the dataset. In total, we have 889 features and 67 (7.5%) of them are sensitive attributes. The same sensitive attribute, such as date of birth, SSN, etc. can appear in multiple datasets but in different formats (for example, 1970-02-31 vs Feb. 1970, 123-45-6789 vs 123456789). The purpose of the experiment is to see whether the proposed pipeline that only learns some of the formats of a certain sensitive attribute, can identify the same attribute in a different format in another dataset. Our idea, as introduced in the methodology section, is to train a model to learn from the metadata. For example, both 1970-02-31 vs Feb. 1970 are date-time, their maximum (and minimum) is all around 2023 (and 1900), etc.

### 3.2 Evaluation of the effectiveness of the machine learning algorithm for de-identification

We randomly split the 889 features into five folds, and we perform five-fold cross validation to evaluate the model's performance. One of the five-fold is selected as the hold-out test set and the remaining four folds for training. We report the mean and variance of the model's performance on the test set, including AUROC, AUPRC, accuracy, prediction, recall, and F1 score.

Following the pipeline in Figure 1, we first feed in the data of each feature to the rule-based sensitive attribute screening module (Algorithm 1) to obtain the probability of each feature being a sensitive attribute. Since this step does not involve any training process, we run the regular expression screening model on all features. In the second step, we run the feature engineering pipeline to calculate the 37 engineered features of all the original features, and then feed them into the machine learning prediction model, which we use gradient boosted trees. The parameters of the model are, number of trees is 100, the maximum tree depth for base learners is 6, learning rate is 0.09.

In Figure 2(a), (b), we demonstrate the performances of using regular expression screening to identify the sensitive attributes by the designed rules. Since the regular repression screening does not involve any training process, the performances in Figure 2(a)(b) are just the screening results on five disjoint folds of the entire dataset. Due to the difference of the sensitive attribute formats, the rules are not guaranteed to always achieve the same accuracy on different sub datasets, and the performance largely depends on how many data samples of a sensitive attribute feature follows matches one of the rules. It is also worth mentioning that, by manually going over all the data sample of all the 899 features, we can always exhaustively list all possible regex patterns to make sure the screening achieves 100% accuracy all the time. However, that is not the purpose of this study, and it is also not a scalable approach in the real world when we face much larger datasets. Nevertheless, an automatic rule learning regex method which can identify unseen patterns and automatically add it to its pattern pool for future usage would be an interesting research direction. It would solve this poor scalability problem and is able to be more accurate as it sees more data. We haven't been able to find a manure toolkit having such a function for the application of detecting PHI healthcare data. Figure 2 (c)(d) shows the performance of the machine learning model using five-fold cross validation on the entire dataset. Our designed features have great generalizability to work for different formats of the sensitive attributes. Of all five folds, the model can identify all sensitive attributes in the test set with both AUROC and AURPC equal to 1.0.

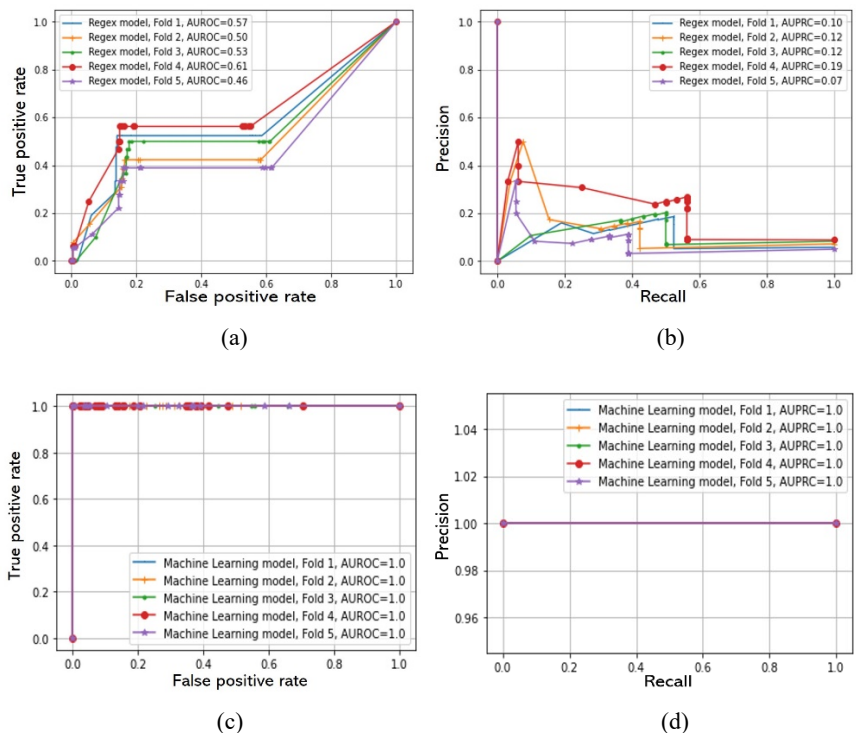

*Figure 2.* Sensitive attribute detection performances (AUROC and AUPRC) on 5-fold cross validation, (a)(b) regular expression screening model, (c)(d) machine learning model

To compare the other performance evaluation metrics, we summarize the cross-fold validation performance of different models in Table 3. The ensemble model's performance is the same as the machine learning prediction model since the latter already achieves perfect performance on our example datasets, i.e., the performance gain provided by the regex model is not present on our datasets. However, it does not serve the reason of neglecting the benefits of regex models, which would very likely improve upon the machine learning model especially the later fails on certain attributes. On the other hand, regular expression has a large running time benefit compared to calculating the engineered features for each attribute. Figure 3 shows the running time of the two methods on 5,000 data samples where regular expression is used to find all the designed patterns for all 5,000 data samples in all columns and ML is used to calculate all the engineered features for all columns. The former only took 1 hour 24 minutes while the latter took 3 hours and 26 minutes, which is near 2.5 times longer, both using single thread. We suggest use multi-threading to compute the engineered feature, for example, for the other experiments (except Figure 3) in this study we used 11 threads (depending on CPU) to reduce the ML method running time to 18.6 minutes.

***Table 3***. *Five-fold cross validation performances of the models on sensitive attribute dentification: mean(std).*

|  | AUROC | AUPRC | sensitivity | specificity | precision | NPV | accuracy | F1 |
|---|---|---|---|---|---|---|---|---|
| **Regex** | .540(.052) | .127(.039) | .479(.064) | .841(.013) | .187(.051) | .956(.008) | .816(.015) | .268(.062) |
| **ML** | 1.0(0.0) | 1.0(0.0) | 1.0(0.0) | 1.0(0.0) | 1.0(0.0) | 1.0(0.0) | 1.0(0.0) | 1.0(0.0) |
| **Ensemble** | 1.0(0.0) | 1.0(0.0) | 1.0(0.0) | 1.0(0.0) | 1.0(0.0) | 1.0(0.0) | 1.0(0.0) | 1.0(0.0) |

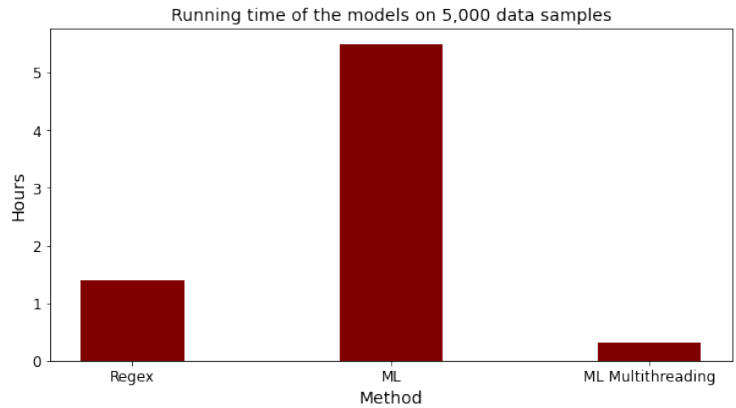

***Figure 3.*** *The running time of the Regex model and the ML model. The running time of regex model is for completely screening all patterns of all 899 columns in the dataset. The running time for ML model is for completely calculating all engineered features of all.*

### 3.3 Explainability

In order to evaluate what engineered features are effective on predicting the sensitive attribute, we use Shapley (SHAP) values to explain the model. SHAP values represent a feature's marginal contribution to the predicted outcome. This is done by evaluating the difference between the prediction of when a feature is included and when it is excluded, while averaging over all possible orderings of the features. In Figure 4, we plot the Top-20 important features ranking by their importance. All values are normalized to the range of 0-1. The error bar stands for standard deviation calculated from the fivefold cross validation. The Gini impurity is the most important feature since it represents the impurity or heterogeneity of a set of labels (so that medical IDs has the largest Gini impurity). The bin counts and bin edge values of histogram is largely dependent on the feature being categorical or continuous Skewness measures the symmetry of the sample distribution and data type such as string, datetime, float, int, etc. also tells whether an attribute likely contains sensitive information. The rest of the features includes statistics around precisions (how many digits are there in the data sample, which works only for numerical features) and the statistics of the feature.

### 4. Discussion

The de-identification of structured data is a crucial step in safeguarding individuals' privacy and ensuring compliance with data protection regulations. This process involves removing identifying information from data sets while preserving their utility. The results of this study have significant implications for this process, emphasizing the need for robust de-identification techniques that can effectively remove identifying information while maintaining the data's usefulness.

However, it is essential to acknowledge the limitations of this study. The sample size was relatively small and may not be representative of the broader population. Additionally, the study relied on data from limited sources, which may be subject to bias and may not generalize to other contexts. Furthermore, the study did not explore certain variables that could have influenced the results, such as participants' socioeconomic status or cultural background.

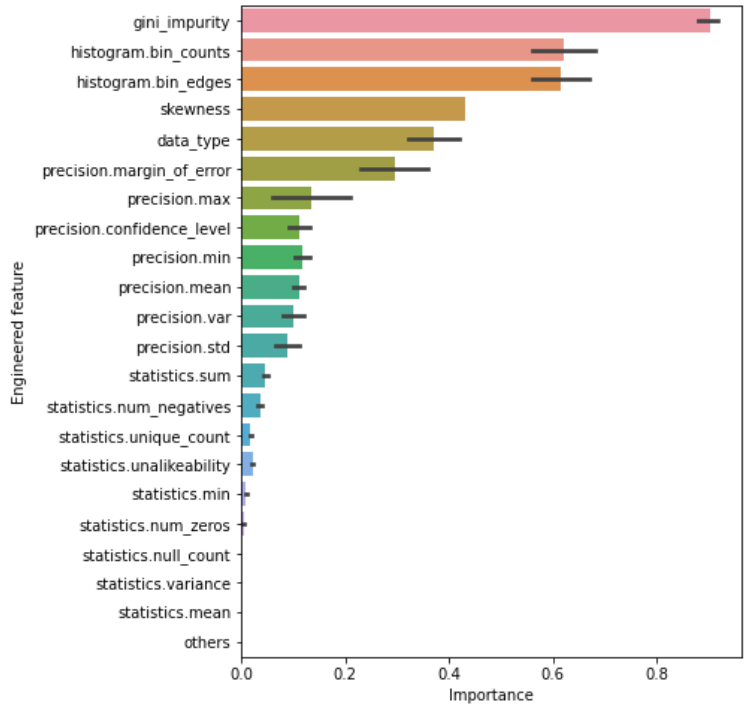

*Figure 4. Feature importance of the engineered features. Importance is normalized to the range of [0,1]. The error bar means standard deviation calculated from five-fold cross validation.*

To address these limitations and expand our understanding of the topic, future research could pursue several avenues. For example, a larger and more diverse sample could be considered to increase the external validity of the findings. Comparative studies across different regions could shed light on the cultural and contextual factors that might influence the results. Additionally, incorporating additional variables or moderators into the analysis could help identify the conditions under which the effects observed in this study are pronounced.

It is also crucial to involve domain experts and stakeholders in the development and evaluation of models to ensure that the de-identification process maintains the data's usefulness and quality for its intended purpose. Regular monitoring and auditing of the models' performance and results are also recommended, with modifications made as necessary to comply with evolving data and privacy regulations.

## 5. Conclusion

In today's data-driven world, the protection of individuals' privacy is of utmost importance. De-identification of structured data is a critical step in ensuring compliance with data protection regulations and safeguarding sensitive information. However, it is important to acknowledge the limitations of current techniques and explore avenues for future research to improve their effectiveness. Determining the appropriate de-identification method and technique requires careful consideration of various factors such as the type of data being used, the desired level of privacy protection, and the applicable regulations. It is crucial to strike a balance between protecting individuals' privacy and maintaining the utility of the data for research and analysis purposes.

One area of particular interest is the use of machine learning in de-identification. While machine learning algorithms have shown promise in automating the de-identification process, there are concerns about their accuracy and potential biases. Therefore, it is important to conduct further research to understand the implications of using machine learning in de-identification and to develop methods to mitigate any potential risks. In light of the limitations of current techniques and the potential of machine learning, it is recommended that future research in this area focuses on developing hybrid approaches that combine the strengths of both human expertise and machine learning algorithms. This approach would involve using machine learning to automatically add rules to rule-based de-identification methods, while also incorporating human oversight to ensure accuracy and fairness.

In conclusion, while de-identification of structured data is a critical step in protecting individuals' privacy and ensuring compliance with data protection regulations, there is still much work to be done to improve the effectiveness of current techniques. By acknowledging the limitations of current methods and pursuing avenues for future research, we can continue to safeguard sensitive information while also advancing research and analysis in various fields.

**Acknowledgement**

XJ is CPRIT Scholar in Cancer Research (RR180012), and he was supported in part by Christopher Sarofim Family Professorship, UT Stars award, UTHealth startup, the National Institute of Health (NIH) under award number R01AG066749, R01LM013712, U54HG012510, OT2OD03270, and U01TR002062, and the National Science Foundation (NSF) #2124789

# Appendix

*Supplementary Table 4*. Sensitive Information Categories and Relevant Regular Expressions

| Sensitive Information | Subcategory | Regular expression patterns | Explanations |
|---|---|---|---|
| **(A) Names** | Firstname, lastame | ^[A-Z]\'?[-a-zA-Z]+$ | match names, A'Bsfs, Absssfs, A-Bsfsfs |
| | Title | (Dr\.|Mr\.|Mrs\.|Ms\.|Miss|Sir|Madam)\s (([A-Z]\'?[A-Z]?[\-a-z]+(\s[A-Z]\'?[A-Z]?[\-a-z]+)*) ) | match salutation |
| | Middle name | \*\*PHI\*\*,? (([A-CE-LN-Z][Rr]?|[DM])\.?) | (([A-CE-LN-Z][Rr]?|[DM])\.?),? \*\*PHI\*\* | match middle initial |
| **(B) Geographic subdivisions** | Postal code | \b(\d{5}(-\d{4})?)\b | match postal code |
| | Address | address_indictor = ['street', 'avenue', 'road', 'boulevard', 'drive', 'trail', 'way', 'lane', 'ave', 'blvd', 'st', 'rd', 'trl', 'wy', 'ln', 'court', 'ct', 'place', 'plc', 'terrace', 'ter', 'highway', 'freeway', 'autoroute', 'autobahn', 'expressway', 'autostrasse', 'autostrada', 'byway', 'auto-estrada', 'motorway', 'avenue', 'boulevard', 'road', 'street', 'alley', 'bay', 'drive', 'gardens', 'gate', 'grove', 'heights', 'highlands', 'lane', 'mews', 'pathway', 'terrace', 'trail', 'vale', 'view', 'walk', 'way', 'close', 'court', 'place', 'cove', 'circle', 'crescent', 'square', 'loop', 'hill', 'causeway', 'canyon', 'parkway', 'esplanade', 'approach', 'parade', 'park', 'plaza', 'promenade', 'quay', 'bypass' 'dr.', 'ave.', 'rd.', 'st.', 'blvd.','pkwy.','city'] | address keywords |
| **(C) Dates** | General date | \b(.*?(?=\b(\d{1,2}[-./\s]\d{1,2}[-./\s]\d{2}|\d{1,2}[-./\s]\d{1,2}[-./\s]\d{4}|\d{2}[-./\s]\d{1,2}[-./\s]\d{1,2}|\d{4}[-./\s]\d{1,2}[-./\s]\d{1,2})\b))\b<br><br>month_name = Jan(uary)?|Feb(ruary)?|Mar(ch)?|Apr(il)?|May|Jun(e)?|Jul(y)?|Aug(ust)?|Sep(tember)?|Oct(ober)?|Nov(ember)?|Dec(ember)? | Date in various formats.<br>YYYY/MM-YYYY/MM<br>MM/YYYY-MM/YYYY<br>MM/YY-MM/YY<br>MM/YY-MM/YY<br>MM/YYYY-MM/YYYY<br>MM/DD-MM/DD<br>DD/MM-DD/MM<br>DD/MM-DD/MM |

| | | | |
|---|---|---|---|
| | | \b(<br>\d{4}[\-/](0?[1-9]\|1[0-2]\|"""+month_name+r""")\-\d{4}[\-/](0?[1-9]\|1[0-2]\|"""+month_name+r""")  # YYYY/MM-YYYY/MM<br>\|(0?[1-9]\|1[0-2]\|"""+month_name+r""")[\-/]\d{4}\-(0?[1-9]\|1[0-2]\|"""+month_name+r""")[\-/]\d{4}  # MM/YYYY-MM/YYYY<br>\|(0?[1-9]\|1[0-2]\|"""+month_name+r""")/\d{2}\-(0?[1-9]\|1[0-2]\|"""+month_name+r""")/\d{2}  # MM/YY-MM/YY<br>\|(0?[1-9]\|1[0-2]\|"""+month_name+r""")/\d{2}\-(0?[1-9]\|1[0-2]\|"""+month_name+r""")/\d{4}  # MM/YYYY-MM/YYYY<br>\|(0?[1-9]\|1[0-2]\|"""+month_name+r""")/([1-2][0-9]\|3[0-1]\|0?[1-9])\-(0?[1-9]\|1[0-2]\|"""+month_name+r""")/([1-2][0-9]\|3[0-1]\|0?[1-9])  #MM/DD-MM/DD<br>\|([1-2][0-9]\|3[0-1]\|0?[1-9])/(0?[1-9]\|1[0-2]\|"""+month_name+r""")\-([1-2][0-9]\|3[0-1]\|0?[1-9])/(0?[1-9]\|1[0-2]\|"""+month_name+r""")  #DD/MM-DD/MM<br>\|(0?[1-9]\|1[0-2]\|"""+month_name+r""")[\-/\s]([1-2][0-9]\|3[0-1]\|0?[1-9])[\-/\s]\d{2}  # MM/DD/YY<br>\|(0?[1-9]\|1[0-2]\|"""+month_name+r""")[\-/\s]([1-2][0-9]\|3[0-1]\|0?[1-9])[\-/\s]\d{4}  # MM/DD/YYYY<br>\|([1-2][0-9]\|3[0-1]\|0?[1-9])[\-/\s](0?[1-9]\|1[0-2]\|"""+month_name+r""")[\-/\s]\d{2}  # DD/MM/YY<br>\|([1-2][0-9]\|3[0-1]\|0?[1-9])[\-/\s](0?[1-9]\|1[0-2]\|"""+month_name+r""")[\-/\s]\d{4}  # DD/MM/YYYY<br>\|\d{2}[\-.\/\s](0?[1-9]\|1[0-2]\|"""+month_name+r""")[\-\.\/\s]([1-2][0-9]\|3[0-1]\|0?[1-9])  # YY/MM/DD<br>\|\d{4}[\-.\/\s](0?[1-9]\|1[0-2]\|"""+month_name+r""")[\-\.\/\s]([1-2][0-9]\|3[0-1]\|0?[1-9])  # YYYY/MM/DD<br>\|\d{4}[\-/](0?[1-9]\|1[0-2]\|"""+month_name+r""")  # YYYY/MM | MM/DD/YY<br>MM/DD/YYYY<br>DD/MM/YY<br>DD/MM/YYYY<br>YY/MM/DD<br>YYYY/MM/DD<br>MM/YYYY<br>MM/YY<br>MM/YYYY<br>MM/DD<br>DD/MM |

| | | `|(0?[1-9]|1[0-2]|"""+month_name+r""")[\-/]\d{4}` # MM/YYYY<br>`|(0?[1-9]|1[0-2]|"""+month_name+r""")/\d{2}` # MM/YY<br>`|(0?[1-9]|1[0-2]|"""+month_name+r""")/\d{2}` # MM/YYYY<br>`|(0?[1-9]|1[0-2]|"""+month_name+r""")/([1-2][0-9]|3[0-1]|0?[1-9])` #MM/DD<br>`|([1-2][0-9]|3[0-1]|0?[1-9])/(0?[1-9]|1[0-2]|"""+month_name+r""")` #DD/MM<br>`)\b` | |
| | DOB | `\b(.*?(?=\b(\d{1,2}[-./\s]\d{1,2}[-./\s]\d{2}` # X/X/XX<br>`|\d{1,2}[-./\s]\d{1,2}[-./\s]\d{4}` # XX/XX/XXXX<br>`|\d{2}[-./\s]\d{1,2}[-./\s]\d{1,2}` # xx/xx/xx<br>`|\d{4}[-./\s]\d{1,2}[-./\s]\d{1,2}` # xxxx/xx/xx<br>`)\b)`<br>`)\b` | Match date of birth |
| | Age | `\b(`<br>`age|year[s-]?\s?old|y.o[.]?`<br>`)\b` | Match age |
| **(D)Telephone/ FAX numbers, (I) Health plan beneficiary numbers, Account numbers, Medical record numbers** | | `\b((\d[\()\-\']?\s?){6}([\()\-\']?\d)+`<br>`|(\d[\()\-.\']?){7}([\()\-.\']?\d)+` # test<br>`)\b`<br>`\b(\d{5}[A-Z0-9]*)\b`<br>`\b([A-Z0-9\-/]{6}[A-Z0-9\-/]*)\b` | Phone/fax/account number/MRN, etc. in the format of SSN/PHONE/FAX XXX-XX-XXXX, XXX-XXX-XXXX, XXX-XXXXXXXX, etc. |
| **(F) Email addresses** | | `\b([a-zA-Z0-9_.+-@\"]+@[a-zA-Z0-9\:\]\[]+[a-zA-Z0-9-.]*)\b` | e.g., xxx.xxx@xxxx.xxx |
| **(G) Social security numbers** | | `\b((\d[\()\-\']?\s?){6}([\()\-\']?\d)+|(\d[\()\-.\']?){7}([\()\-.\']?\d)+)\b` | e.g., xxx-xx-xxxx, etc. |
| **(N) Web URLs** | | `\b((http[s]?://)?([a-zA-Z0-9$-_@.&+:!*\(\),])*[\.\/]([a-zA-Z0-9$-` | e.g., http://xxx.xxx.xxx/xxx.xxx |

| | | _@.&+:\!\*\(\),])*)\b | |
|---|---|---|---|
| **(O) Internet Protocol (IP) addresses** | | ^(?:(?:25[0-5]\|2[0-4][0-9]\|[01]?[0-9][0-9]?)\.){3}(?:25[0-5]\|2[0-4][0-9]\|[01]?[0-9][0-9]?)$ | e.g., xxx.xxx.xxx.xxx |
| **(R) Other unique identifiers** | Race | "(White/Caucasian\|White\|Caucasian\|American\|Indian\|Alaska\|Native\|Asian\|Black\|African\|Native\|Hawaiian\|Pacific\|Islander)" | Match race |
| | Gender | "(Male\|Female)" | Match gender |
| | Ethnicity | "(Hispanic\|Latino)" | Match ethnicity |

**Supplementary Table 5**. *Summary of the engineered features*

| Engineered features | Description |
|---|---|
| Data_type | The data type of this column: int, float, string, datetime |
| categorical | Whether the data is categorical data type |
| order | whether the data in this column is ordered |
| null_count | the number of null entries in this column |
| min | minimum value in the sample |
| max | maximum value in the sample |
| mode | mode of the entries in the sample |
| median | median of the entries in the sample |
| median_absolute_deviation | the median absolute deviation of the entries in the sample |
| sum | the total of all sampled values from the column |
| mean | the average of all entries in the sample |
| variance | the variance of all entries in the sample |
| stddev | the standard deviation of all entries in the sample |
| skewness | the statistical skewness of all entries in the sample |
| kurtosis | the statistical kurtosis of all entries in the sample |
| num_zeros | the number of entries in this sample that have the value 0 |
| num_negatives | the number of entries in this sample that have a value less than 0 |
| histogram | contains histogram relevant information |
| bin_counts | the number of entries within each bin |
| bin_edges | the thresholds of each bin |
| quantiles | the value at each percentile in the order they are listed based on the entries in the sample |
| categories | a list of each distinct category within the sample if categorical = 'true' |
| unique_count | the number of distinct entries in the sample |
| unique_ratio | the proportion of the number of distinct entries in the sample to the total number of entries in the sample |
| categorical_count | number of entries sampled for each category if categorical = 'true' |
| gini_impurity | measure of how often a randomly chosen element from the set would be incorrectly labeled if it was randomly labeled according to the distribution of labels in the subset. Gini_impurity $=\sum_{i=1}^{K} p_i(1 - p_i)$, where $k$ is the number of categories in the column, and $p_i$ is the proportion of the samples that belong to category $i$. |
| diversity index | A value denoting how frequently entries differ from one another within the sample |
| precision | A dict of statistics with respect to the number of digits in a number for each sample |

**Engineered features designing logics**

*Data type:* sensitive attributes can be more often to have certain data types than non-sensitive attributes, such as date of birth (datetime), address (string). On the other hand, non-sensitive attributes such as lab tests, vitals, medication dosages are often in float format.

*Categorical attributes:* sensitive attributes are often categorical, such as sex, ethnicity.

*Order*: if the values of a column follows a certain order, it can be a sensitive attribute such as medical record numbers, encounter IDs. This depends on whether the data samples have been sorted or not.

*Null count*: columns with a lot of null values are more often to be non-sensitive attributes, especially for the lab tests and medication columns when not all patients have performed the test or been prescribed.

*Min*: The minimum value of a column. Too large a minimum value of a column may indicate this is a sensitive attribute, such as medical record number; whereas non-sensitive attributes (lab test result, medication dosage) minimum number can be small.

*Max/Median/median absolute deviation/sum/mean*: Same logic as the minimum number.

*Mode*: If there is no mode for the data, all data samples are unique and it is highly likely to be a sensitive attribute. Need to be combined with other features to decide, because for example, lab tests results in float numbers are very unlikely to have two of the same numbers among a certain number of data samples.

*Variance/Standard deviation*: Unique values have very large variance and lab tests, vitals values are relatively more concentrate

*Skewness*: For example, sensitive attributes like medical record numbers and encounter IDs are likely to have a symmetric distribution, with an equal number of values on either side of the mean, resulting in a skewness value of 0. On the other hand, non-sensitive attributes like lab test results and medication dosages are likely to have a skewed distribution, where the tail is longer on one side of the mean. This can be due to factors such as outliers or a natural skewness in the underlying biological processes.

*Kurtosis*: Non-sensitie attributes such as lab tests, vitals signs can follow a normal distribution (most data falls beside the mean and it becomes less as data being further away from the mean) has a kurtosis of 3, which is considered as the reference value for kurtosis. This means that the normal distribution has neither peakedness nor flatness compared to other distributions. On the other hand, sensitive attribute such as medical IDs has a uniform distribution, with a kurtosis of -1.2. This indicates that the uniform distribution has flatter peaks and thinner tails compared to the normal distribution.

*Num zeros/Num negatives:* medications, diagnosis, treatment often uses 0 to indicate not prescribed with a medication, not diagnosed of a disease or not receiving a treatment

*Histogram (bin counts, bin edges)*: For sensitive attributes, we may expect to see a more uniform distribution of values across the bins or a few dominant values with a small number of other values scattered across the bins. For example, if we look at a histogram of birth year, we may expect to see a relatively uniform distribution with a slight peak for the most common birth year. On the other hand, non-sensitive attributes like lab results may have more skewed distributions with a few dominant values and a large number of values clustered around zero or a specific range.

*Bin counts*: For example, if a column has a small number of distinct values with a high frequency in each bin, it may indicate a sensitive attribute, such as a patient's unique medical record number or social security number. On the other hand, non-sensitive attributes such as lab test results or medication dosages may have a larger number of distinct values with more evenly distributed bin counts.

*Bin edges*: For example, if a column has a narrow range of values with a small number of distinct values, it may be a sensitive attribute. On the other hand, if the column has a wide range of values with many distinct values, it is less likely to be a sensitive attribute. Additionally, if the bin edges reveal patterns or clusters of values, it may indicate that the column contains sensitive information.

*Quantiles (0.25, 0.75)*: first quartile (Q1) and third quartile (Q3) can provide insights into the distribution of the values in the column. For sensitive attributes, the values may be more concentrated around a certain range, resulting in a smaller interquartile range (IQR) and a higher median. In contrast, for non-sensitive attributes, the values may be more spread out, resulting in a larger IQR and a lower median.

*Categories*: if a column contains sensitive attributes such as ethnicity, race, or religion, then the values in the column will be unique and categorical. On the other hand, non-sensitive attributes such as lab tests, medication dosages, and vitals, will have a wide range of values and are not categorical.

*Unique count:* Sensitive attributes may have a smaller number of distinct entries than non-sensitive attributes, especially for categorical attributes like sex and ethnicity.

*Unique ratio:* if a column has a high proportion of distinct entries (i.e., a large number of unique values compared to the total number of entries in the column), it could suggest that the column contains sensitive information that is unique to each individual.

*Categorical count:* helps to identify the distribution of the data across different categories, and can provide insights into whether a particular attribute is sensitive or not.

*Gini impurity:* the probability of a data sample is labeled wrong if it is randomly labeled according to the distributions of labels in this column

*Diversity index:* If a column has high diversity, meaning that the values in that column are unique or distinct from one another, it may indicate that the column contains sensitive information, such as a medical record number or an encounter ID.

*Precision (min, max, mean, var, std, sample size, margin of error, confidence level)*: Statistics with respect to the number of digits in a number for each sample, because sensitive attributes may have a uniform number of digits, e.g., medical record number.